\documentclass[11pt]{article}
\usepackage{amsmath}
\usepackage{amsfonts,amssymb}

\begin{document}

\newcommand{\nl}{\nonumber\\}
\newcommand{\nnl}{\nl[6mm]}
\newcommand{\nle}{\nl[-2.5mm]\\[-2.5mm]}
\newcommand{\nlb}[1]{\nl[-2.0mm]\label{#1}\\[-2.0mm]}
\newcommand{\ab}{\allowbreak}

\renewcommand{\leq}{\leqslant}
\renewcommand{\geq}{\geqslant}

\newcommand{\be}{\bes}
\newcommand{\ee}{\ees}
\newcommand{\bes}{\begin{eqnarray}}
\newcommand{\ees}{\end{eqnarray}}
\newcommand{\eens}{\nonumber\end{eqnarray}}

\renewcommand{\/}{\over}
\renewcommand{\d}{\partial}
\newcommand{\e}{{\rm e}}

\newcommand{\ket}[1]{\big{|}#1\big{\rangle}}
\newcommand{\vac}{\ket{\hbox{vac}}}

\newcommand{\dlt}{\delta}

\newcommand{\dqids}{{{d q^i(s)\/d s}}}
\newcommand{\dqkds}{{{d q^k(s)\/d s}}}

\newcommand{\no}[1]{{\,:\kern-0.7mm #1\kern-1.2mm:\,}}

\newcommand{\vect}{{\mathfrak{vect}}}

\renewcommand{\L}{{\cal L}}
\newcommand{\Lxi}{\L_\xi}

\newcommand{\ZZ}{{\mathbb Z}}

\title{{A diffeomorphism anomaly in every dimension}}

\author{T. A. Larsson \\
Vanadisv\"agen 29, S-113 23 Stockholm, Sweden\\
email: t.larsson@hdd.se}

\maketitle 
\begin{abstract} 
Field-theoretic pure gravitational anomalies only exist in $4k+2$
dimensions. However, canonical quantization of non-field-theoretic
systems may give rise to diffeomorphism anomalies in any number of
dimensions. I present a simple example, where a higher-dimensional
generalization of the Virasoro algebra arises upon quantization.
\end{abstract}

The Polyakov action in string theory is nothing but 1+1-dimensional
gravity coupled to scalar fields \cite{GSW87}. It thus seems reasonable
to expect that canonical quantization of 3+1-dimensional gravity
should proceed along similar lines. We know from string theory that
quantization of the coordinate fields gives rise to a conformal
anomaly, eventually to be cancelled by ghosts. Analogously, one may
expect that quantization of the fields alone gives rise to an anomaly
in the diffeomorphism constraint.

At first sight, this idea appears to be ruled out by two no-go 
theorems:
\begin{enumerate}
\item
The diffeomorphism algebra in $N$ dimensions has no central extension,
except when $N=1$ \cite{RSS89}.
\item
There are no pure gravitational anomalies in 3+1 dimensions.
\end{enumerate}
However, these no-go theorems can be evaded. 
The extension (\ref{Vir}) that generalizes the Virasoro algebra to
several dimensions is not a central one, except in one dimension.
In general, it is an extension of $\vect(N)$, the algebra of vector 
fields (diffeomorphism algebra) in $N$ dimensions, by its modules 
of closed $(N-1)$-forms.
A closed zero-form is a constant function is the trivial
module, so the Virasoro extension is central when $N=1$, but not 
otherwise.

The claim that there are no diffeomorphism anomalies in four
dimensions is simply false. What is true and has been proven is that
there are no {\em field-theoretic} diffeomorphism anomalies
\cite{Bon86}. Non-field-theoretic diffeomorphism anomalies can and do
arise. In this note I present a simple system which gives rise to an
extension of the diffeomorphism algebra in $N$ dimensions upon
canonical quantizion.

Consider a $2N$-dimensional phase space with coordinates $q^i$ and
$p_i$, $i = 1,2,...,N$, and Poisson brackets 
\be 
\{q^i, p_j\} = \dlt^i_j, \qquad \{p_i, p_j\} = \{q^i, q^j\} = 0.
\ee 
$\vect(N)$ is generated by vector fields of the
form $\xi = \xi^i(x)\d_i$, where $\d_i = \d/\d x^i$. It evidently
acts on the space of functions of this phase space.
Upon canonical quantization, the Poisson brackets are replaced by 
commutators
\be 
[q^i, p_j] = i\dlt^i_j, \qquad [p_i, p_j] = [q^i, q^j] = 0.
\label{pq1} 
\ee 
$q^i$ and $p_j = -i\d/\d q^j$ become
operators acting on the space of functions depending on the $q$'s
alone, and $\vect(N)$ acts on this space as the operator 
$\Lxi = i\xi^i(q) p_i$. This
expression is already normal ordered, and there is no anomaly. In
fact, it is easy to verify that 
$[\Lxi, \L_\eta] = \L_{[\xi,\eta]}$, 
where $[\xi,\eta] = \xi^i\d_i\eta^j\d_j - \eta^j\d_j\xi^i\d_i$.

There is a  simple generalization of this system in
which $\vect(N)$ does acquire an extension. Consider the 
infinite-dimensional phase space with coordinates $q^i(s)$
and $p_i(s)$, where $s \in S^1$ is a parameter on a circle.
We quantize this system by promoting $q^i(s)$ and  $p_i(s)$
to operators satisfying the canonical commutator relations
\be
[q^i(s), p_j(s')] = i\dlt^i_j \dlt(s-s'), \qquad
[p_i(s), p_j(s')] = [q^i(s), q^j(s')] = 0.
\label{pq2}
\ee
The Hilbert space is some space of functionals of half
of the oscillators. One possibility is to consider the space of
functionals of $q^i(s)$, with $p_i(s) = -i\dlt/\dlt q^i(s)$. 
In other words, we define a vacuum $\vac$ by $p_i(s)\vac = 0$,
so $q^i(s)$ are creation operators and $p_i(s)$ are annihilation
operators. $\vect(N)$ acts on this space without anomaly.

However, there is no Stone-von Neumann theorem in infinite dimension,
so the choice of vacuum is important. A different choice does lead
to an anomaly. Expand the phase space functions in a Fourier series
in $s$, i.e.
\bes
q^i(s) = \sum_{k=-\infty}^\infty \hat q^i(k) \e^{-iks} \equiv
q^i_<(s) + \hat q^i(0) + q^i_>(s), \nle
p_i(s) = \sum_{k=-\infty}^\infty \hat p_i(k) \e^{-iks} \equiv
p_i^<(s) + \hat p_i(0) + p_i^>(s).
\eens
where $q^i_<(s)$ is the sum over negative Fourier ($k<0$) modes 
alone, etc.
Now define the new vacuum $\ket0$ by requiring that $q^i_<(s)$ and
$p_i^\leq(s) = p_i^<(s) + \hat p_i(0)$ annihilate it. 
Normal ordering is defined as usual by moving the annihilation
operators to the right. 
The normal-ordered $\vect(N)$ generators read
\bes
\Lxi &=& i\int ds\ \no{\xi^i(q(s)) p_i(s)} 
\nle
&\equiv& i\int ds\ ( \xi^i(q(s)) p_i^\leq(s) 
 +  p_i^>(s) \xi^i(q(s)) ).
\eens
By a straightforward calculation one finds that $\vect(N)$
has acquired an extension which generalizes the Virasoro algebra
to $N$ dimensions \cite{Lar98}:
\bes
[\Lxi,\L_\eta] &=& \L_{[\xi,\eta]} 
 + {1\/2\pi i}\int ds\ \dqkds  
 \Big\{ c_1 \d_k\d_j\xi^i(q(s))\d_i\eta^j(q(s)) + 
\nlb{mVir}
&&\quad+ c_2 \d_k\d_i\xi^i(q(s))\d_j\eta^j(q(s)) \Big\},
\eens
where $c_1 = 1$ and $c_2 = 0$. The classical system defined by 
(\ref{pq2}) has thus acquired a diffeomorphism 
anomaly upon quantization.

This construction of anomalous $\vect(N)$ representations, which was
first discovered in \cite{RM94}, can be considerably generalized. In
fact, one can construct representations labelled by a $gl(N)$
representation and a positive integer \cite{Lar98}. The constant $c_2
\neq 0$ for some of these systems.

We now proceed to construct a Hamiltonian which commutes with
(\ref{mVir}).
\bes
H  &=& \int ds\ \no{\dqids p_i(s)}  \nle
&\equiv& \int ds\ (\dqids p_i^\leq(s) + p_i^>(s)\dqids)
\label{Ham}
\eens
is bounded from below, because the creation operators have 
positive eigenvalues:
\be
[H, \hat q^i(k)] = k \hat q^i(k), \qquad
[H, \hat p_i(k)] = k \hat p_i(k).
\ee
Moreover, $[\Lxi, H] = 0$, so the
anomalous diffeomorphism algebra (\ref{mVir}) is a symmetry of this
Hamiltonian.

The equations of motion are somewhat pathological. The equations of
motion for $q^i(s,t)$ and $p_j(s,t)$ read
\be
{\d q^i\/\d t} = {\d q^i\/\d s}, \qquad
{\d p_j\/\d t} = {\d p_j\/\d s}, 
\ee
with the solution $q^i(s,t) = q^i(s+t)$, $p_j(s,t) = p_j(s+t)$.
In particular, the non-zero Poisson brackets at non-equal times
become
\be
\{q^i(s,t), p_j(s',t')\} = \dlt^i_j \dlt(s+t-s'-t').
\ee

We note that the Hamiltonian can be embedded into the family of
operators
\be
H_k = \int ds\ \e^{iks} \no{\dqids p_i(s)},
\ee
which generate an additional Virasoro algebra. Classically, this
algebra commutes with $\vect(N)$, but normal ordering introduces an
anomaly \cite{Lar98}. However, $\Lxi$ still commutes with the
Hamiltonian $H = H_0$.
We obtain a lowest-weight representation of this extra Virasoro
algebra, since $H_k\ket0 = 0$ for all $k < 0$.

To make the analogy with the Virasoro algebra very explicit, it is
instructive to use a Fourier basis on the $N$-dimensional torus
with coordinates $x = (x^i)$. Let $L_i(m)$ be the generator
$\Lxi$ corresponding to the vector field 
$\xi = -i \exp(i m_k x^k)\d_i$,
where $m = (m_i) \in \ZZ^N$ labels momenta. Moreover, set
\be
S^i(m) = {1\/2\pi i}\int ds\ \dqids \exp(im_kq^k(s)).
\ee
The multi-dimensional Virasoro algebra (\ref{mVir}) takes the 
form \cite{Lar91,RM94}
\bes
[L_i(m), L_j(n)] &=& n_i L_j(m+n) - m_j L_i(m+n) \nl
&&  + (c_1 m_j n_i + c_2 m_i n_j) m_k S^k(m+n), \nl
{[}L_i(m), S^j(n)] &=& n_i S^j(m+n)
+ \delta^j_i m_k S^k(m+n), 
\label{Vir}\\
{[}S^i(m), S^j(n)] &=& 0, \nl
m_i S^i(m) &=& 0.
\eens
It is easy to see that (\ref{Vir}) generalizes the Virasoro
algebra to higher dimensions. Namely, in one dimension we can
ignore all indices, and the unique solution of the closedness
condition $m S(m) = 0$ is proportional to a Kronecker delta, 
$S(m) \propto \delta(m)$. If we make this substitution into the
other equations of (\ref{Vir}), we immediately recover the defining
relations for the Virasoro algebras, with the trivial linear 
cocycle put to zero. In particular, $[L(m), S(n)] = 0$.

The diffeomorphism anomalies described in this note are very different
from field- and string-theoretical gravitational anomalies. In string
theory, diffeomorphism anomalies arise from chiral fermions and only
exist if spacetime has $4k+2$ dimensions (\cite{GSW87}, part II, page
336). Such anomalies, like pure gauge anomalies in the standard model,
are very problematic from a mathematical point of view; in the
Hamiltonian formalism, they give rise to Mickelsson-Faddeev algebras
which apparently lack good representations \cite{Pic89}. In contrast,
in this note we have constructed diffeomorphism anomalies that are
similar to the conformal anomaly in the bosonic string. The
quantization of the fields alone (without ghosts) gives rise to
lowest-weight representations of an extension of the classical
symmetry algebra (the Virasoro algebra in one or several dimensions),
which has nothing to do with chiral fermions.

This expectation can be made mathematically rigorous. It is well known
that all non-trivial unitary lowest-weight representations of the
Virasoro algebra have a positive value of the central charge (discrete
unitary series and $c \geq 1$). A unitary representation in higher
dimensions must give rise to a unitary representation of each Virasoro
subalgebra generated by vector fields depending on a single variable.
If the representation of the higher-dimensional Virasoro algebra is
non-trivial, at least one such subalgebra must be non-trivially
represented, i.e. there must be an anomaly. This proves that if
quantization of 3+1-dimensional gravity involves non-trivial, unitary,
lowest-weight representations of the diffeomorphism constraint, as one
may expect on general grounds, a diffeomorphism anomaly of the kind
described above must arise.

Thiemann has recently quantized the Nambu-Goto string using methods
from Loop Quantum Gravity (LQG) \cite{Thi04}. This construction has
been criticized on the grounds that the conventional conformal anomaly
does not appear, because the representations in LQG, albeit unitary,
are not of lowest-weight type. More precisely, one can argue that LQG
is not canonical quantization in the conventional sense, which should
give rise to a conformal anomaly in the bosonic string and to a
diffeomorphism anomaly in 3+1-dimensional gravity, eventually to be
cancelled by ghosts. What has not been widely appreciated is that 
the same critique applies to string theory, which
is also incapable of producing the necessary diffeomorphism anomaly,
at least in four dimensions.

To conclude, I have described a classical canonical system whose
Hamiltonian commutes with diffeomorphisms, and where the
diffeomorphism algebra acquires non-trivial quantum corrections. One
of the main open problems is how to define unitarity; since there is
no invariant inner product, only dual spaces can be invariantly
paired. Nevertheless, the fact that diffeomorphism anomalies in every
dimension are possible outside a field-theoretical framework is quite
striking and apparently not widely known.

\end{document}